\documentclass[aps,prd,twocolumn,showpacs,amsmath,nofootinbib]{revtex4-1}

\usepackage{amsmath}
\usepackage{amssymb}
\usepackage{amsfonts}
\usepackage{mathrsfs}
\usepackage{bm}      
\usepackage{slashed}
\usepackage{latexsym}
\usepackage{enumitem}

\usepackage[T1]{fontenc}
\usepackage{lmodern}
\usepackage{anyfontsize}

\usepackage{threeparttable}
\usepackage{color}
\usepackage{hyphenat}
\usepackage{float}
\usepackage{graphicx}
\usepackage{indentfirst}
\usepackage{setspace}
\usepackage{ragged2e}

\usepackage{booktabs}
\usepackage{multirow}
\usepackage{tabularx}
\usepackage{makecell}
\usepackage{dcolumn}
\usepackage{adjustbox}
\usepackage{subcaption}

\usepackage{microtype}

\usepackage[colorlinks,
            linkcolor=blue,
            citecolor=blue,
            urlcolor=blue,
            breaklinks=true]{hyperref}

\begin{document}

\title{Galactic constraints on spinning primordial black hole dark matter from interstellar dust heating}

\author{Shaobin Hu}
\author{Yupeng Yang}\email{ypyang@aliyun.com}
\author{Chengjie Sun}
\author{Zihan Li}
\author{Jiafan Sun}
\author{Yuzhu Tong}
\author{Yankun Qu}
\author{Shuangxi Yi}

\affiliation{ School of Physics and Physical Engineering, Qufu Normal University, Qufu, Shandong, 273165, China }

\begin{abstract}
Primordial black holes (PBHs) are compelling dark matter candidates. PBHs with masses between $10^{15}$ and $10^{18}\,\mathrm{g}$ can heat interstellar dust via Hawking radiation. Previous studies of this dust heating mechanism mostly neglected PBH spin and adopted a single dark matter halo profile. In this work, we incorporate PBH spin, which substantially enhances the emitted radiation flux, and systematically investigate the dependence of constraints on the dark matter density distribution by considering five different halo models. We compute the complete photon spectra, including both primary and secondary emissions. Our results show that, for a fixed profile and mass function, larger spin parameters yield stronger constraints on the PBH fraction $f_{\mathrm{PBH}}$. Among the halo models, the Isothermal profile gives the most stringent limits, followed by Einasto, then NFW and Moore, while the Burkert profile yields the weakest constraints. For silicate grains, which cool less efficiently than graphite, the upper limits reach $\mathcal{O}(10^{-4})$ for high spin cases. We consider both monochromatic and lognormal mass functions, and find consistent trends between them. For the lognormal case, larger values of the width $\sigma$ lead to a broader mass range being excluded, in particular ruling out massive PBHs as the sole dark matter component. Our bounds are generally weaker than other existing limits, but they provide a complementary and independent constraint.
\end{abstract}

\maketitle

\section{Introduction}
The existence of dark matter has been firmly established through a wealth of astronomical and cosmological observations \cite{2026Univ...12...48C,2006ApJ...648L.109C,2004ApJ...604..596C}. One of its key signatures is that the rotation curves of galaxies at large radii do not decline as expected from visible matter alone, but instead remain approximately flat. Dark matter also induces a more pronounced gravitational lensing effect. Furthermore, it plays a crucial role in galaxy cluster collisions and in shaping the large‑scale structure of the Universe \cite{2017IJMPD..2630012F,2024JCAP...10..075L,1985ApJ...292..371D}. Many dark matter candidates have been proposed, such as, sterile neutrinos, axions, and weakly interacting massive particles (WIMPs) \cite{2016PhRvL.117f1101S,2018RPPh...81f6201R,Roszkowski:2004jc,Schumann:2019eaa,Bertone:2004pz,Green:2021jrr,Jungman:1995df}. Primordial black holes (PBHs) are also considered as dark matter candidates \cite{2016PhRvL.116t1301B,2024NuPhB100316494G,2017PDU....15..142C}. Current observational constraints indicate that only PBHs within specific mass ranges can account for a fraction of dark matter \cite{2020ARNPS..70..355C,2021arXiv211002821C}. Therefore, investigating the properties of PBHs is of considerable significance.

To investigate the properties of PBHs, various observational and theoretical constraints have been derived from a wide range of sources~\cite{2019PhRvL.123y1102D,2020PhRvL.125j1101D,2018PhRvD..98d3006C,2023PhRvD.107f3535M,2010PhRvD..81j4019C,2019PhRvL.122d1104B,2017PhRvD..96b3514C,2020PhRvD.101l3514L,2018PhRvL.121n1101Z,2019PhRvD..99h3503N,2010PThPh.123..867S,2021JCAP...04..062D,2024EPJC...84..606S,Yang:2025mkp,Hao:2024hzu,Yang:2023qnl,Yang:2022puh,Yang:2022nlt,Yang:2021agk,Yang:2021idt,Yang:2020egn,Yang:2020zcu,Yang:2026zcq,Li:2026paw,Khan:2025kag,Mittal:2021egv,Ray:2021mxu,Balaji:2025afr}. Recently, the authors of Ref.~\cite{2023PhRvD.107f3535M} investigated the heating of interstellar dust by PBHs and derived constraints on the PBH abundance using the temperature of dust in the Galaxy. However, their analysis was restricted to non-spinning PBHs and adopted a single dark matter halo profile, the Navarro-Frenk-White (NFW) model. In this work, we extend their study in several significant ways. First, we incorporate the spin of PBHs into our calculations, which is a crucial factor as it can substantially enhance the Hawking radiation flux, leading to a more pronounced dust heating effect. Second, we go beyond the NFW profile by systematically examining four additional dark matter halo models to assess the model dependence of the resulting constraints. Furthermore, we also evaluate the contribution of electrons emitted by PBHs to the dust heating, providing a more comprehensive analysis.
       
The structure of this paper is as follows. In Sect.~\ref{sec:basic}, we introduce the basic properties of PBHs and interstellar dust. Sect.~\ref{sec:heat_from_pbh} presents the heating mechanism of dust by PBHs, and the derived constraints on PBHs are given in Sect.~\ref{sec:constraints}. Finally, we summarize our conclusions in Sect.~\ref{sec:con}.

\section{The basic characteristics of primordial black hole and interstellar dust}
\label{sec:basic}

\subsection{The primordial black hole}

Primordial black holes (PBHs) are a special class of black holes that are theoretically formed in the very early Universe. They are fundamentally distinct from stellar‑mass black holes, which arise from the gravitational collapse of massive stars. Following cosmic inflation, during the radiation‑dominated epoch, quantum fluctuations in the early Universe generated significant density perturbations. When local overdensities exceeded a critical threshold, gravitational collapse proceeded rapidly in these regions, ultimately giving rise to PBHs \cite{2017JCAP...09..013G,2021PhLB..81936468M}. These objects formed at extremely early cosmic times, long before the emergence of stars and galaxies, with formation epochs potentially as early as \(\sim 10^{-23}\) seconds after the Big Bang \cite{2010PhRvD..81j4019C}.

      The mass range of PBHs spans an enormous interval, from $\sim 10^{-5}\,\mathrm{g}$ to several thousand solar masses. However, according to theory, PBHs with masses below 
$\sim 10^{15}\,\mathrm{g}$ would have completely evaporated via Hawking radiation over the age of the Universe. For PBHs with masses greater than $\sim 10^{18} \rm g$, the Hawking temperature is too low for any appreciable radiation to occur. Consequently, the influence of such black holes on the Universe is primarily gravitational, manifesting through microlensing effects and gravitational wave signatures \cite{2024ApJ...976L..19M,2021arXiv211106990K,2025PDU....5002072H,2026JCAP...02..063W,2025PDU....4901991D,2025PDU....4901991D,2025arXiv250707665Y,Mohammadi:2025avz}. 

Primordial black holes (PBHs) are among the most important candidates for dark matter. This is because PBHs could exist in extremely large numbers and interact with other matter almost exclusively through gravity, which aligns closely with the core properties of dark matter. Interest in PBHs as dark matter candidates has grown significantly following the first LIGO detections of gravitational waves~\cite{2016PhRvL.116t1301B,2017JCAP...09..037R,2017JCAP...09..013G}. However, current observational constraints indicate that only PBHs within specific mass ranges can contribute to dark matter~\cite{2021arXiv211002821C}. For example, certain intermediate mass intervals have been ruled out as dominant dark matter components owing to the lack of observed gravitational lensing events. Furthermore, PBH binaries in the early Universe can convert mass into gravitational waves \cite{2019JCAP...02..018R,2017JCAP...09..037R}. Although this effect is suppressed, it remains physically viable.

In general, many theories show that the spin of PBHs is highly dependent on their formation mechanisms and the surrounding cosmological environment. In the standard scenario where PBHs form from the collapse of rare overdense peaks in a radiation-dominated universe, the angular momentum is severely suppressed~\cite{Chiba:2017rvs,Mirbabayi:2019uph}. In this case, the spin parameter is typically small, arising only at second order in the primordial density perturbations, with a dimensionless spin parameter $a_* \lesssim 0.4$ and often at the percent level~\cite{DeLuca:2019buf}, as the collapse of near-spherical high-density peaks inhibits significant rotation. However, alternative formation channels can yield PBHs with considerable or even near-extremal spins. For instance, during a prolonged matter-dominated era, the growth of angular momentum prior to collapse can lead to rapid rotation with $a_{\star}\simeq 1$~\cite{1980PhLB...97..383K,2017PhRvD..96h3517H,He:2019cdb}. Similarly, PBHs produced through primordial phase transitions~\cite{Cotner:2018vug,Cotner:2019ykd,Wang:2025hwc}, scenarios involving a stiff equation of state~\cite{Kokubu:2018fxy}, or other exotic mechanisms~\cite{Bai:2019zcd,Arbey:2019jmj,Dvali:2021byy} may also acquire substantial spin. 

The temperature of an uncharged rotating PBH with mass $M_{\rm PBH}$ can be written as~\cite{PhysRevD.13.198,PhysRevD.14.3260}

\begin{equation}
T_{\rm PBH}=\frac{1}{4\pi G_{N} M_{\rm PBH}}\frac{\sqrt{1-a^{2}_{*}}}{1+\sqrt{1-a^{2}_{*}}},
\end{equation}
where $G_{N}$ is the gravitational constant, and $a_{*}=J/(G_{N}M^{2}_{\rm PBH})$ is the spin parameter, with $J$ the angular momentum of the PBH. 
The particle number $N$ emitted from a PBH with spin $s$ per unit energy and time is 

\begin{equation}
\frac{d^{2}N}{dEdt}=\frac{1}{2\pi}\frac{{\rm \Gamma}_{s}(E,M_{\rm PBH},a_{*})}{{\rm exp}(E/T_{\rm PBH})-(-1)^{2s}},
\label{eq:energy_spectrum}
\end{equation}
where ${\rm \Gamma}_{s}$ is the graybody factor~\cite{PhysRevD.13.198,PhysRevD.14.3260,PhysRevD.16.2402}. Including spin, a PBH will 
emit more particles compared with a non-spinning one~\cite{2020PhRvD.101b3010A,2024PhRvD.110l3022L,PhysRevD.44.376,2020PhRvL.125j1101D,Arbey:2019mbc,Arbey:2026koc}. Since the PBH temperature scales inversely with mass as $T_{\rm PBH}\sim M^{-1}_{\rm PBH}$, and PBHs with masses below $\sim 10^{15}\rm g$ would have completely evaporated by the present epoch, we focus on the mass range 
$10^{15}-10^{17}\rm g$ in this work. To compute the energy spectrum in Eq.~(\ref{eq:energy_spectrum}), we employ the public code \texttt{BlackHawk}~\cite{Arbey:2019mbc,Arbey:2026koc,Calza:2025whq,Arbey:2021mbl}.

\subsection{The interstellar dust}

Interstellar dust is a critical component of the interstellar medium, with a mass about 1\% of that of interstellar gas. Its composition is commonly modeled as a mixture of amorphous silicates and carbonaceous (graphitic) grains, along with polycyclic aromatic hydrocarbons (PAHs)~\cite{Draine:2003if}. The grain size distribution typically follows a power law $n(a) \propto a^{-3.5}$, corresponding to the Mathis-Rumpel-Nordsieck (MRN) model, 
with radii ranging from $\sim 0.005$ to 2.5 $\mu \rm m$~\cite{1977ApJ...217..425M}. Dust grains efficiently absorb and scatter starlight, producing a characteristic wavelength-dependent extinction curve that exhibits a prominent absorption bump at 2175 $\AA$, attributed to graphitic or PAH components~\cite{Draine:2006dn}. The absorbed stellar energy is re-radiated in the infrared to submillimeter regimes, with PAHs responsible for strong emission features at 3.3, 6.2, 7.7, 8.6, and 11.3 $\mu \rm m$~\cite{Draine:2006dn}. 

The thermal behavior of dust depends strongly on grain size. Large grains ($a\gtrsim 0.01 \mu \rm m$) maintain equilibrium temperatures of 10-20 K in diffuse clouds, 30-200 K in HII regions, and up to 1000-1500 K in circumstellar envelopes
~\cite{1981ApJ...248..138D,1986ApJ...302..363D,Draine:2000dj,Draine:2006dn,Ivlev:2015zsa,Kalvans_2018,2005pcim.book.....T}. Small grains, however, have low heat capacities and experience stochastic heating by individual starlight photons, resulting in transient temperature spikes; between such spikes, they cool down to the cosmic microwave background temperature of 2.7 K~\cite{Horn:2007kz,Draine:2000dj}. Dust grains become electrically charged through electron attachment and photoionization, enabling interactions with electromagnetic fields. This charging process, along with collisional plasma collection and cosmic-ray-induced photoemission, significantly influences grain coagulation, surface chemistry, and the dynamical coupling between gas and magnetic fields in molecular clouds~\cite{Ivlev:2015zsa,2013A&A...556A...6B}. In this work, we neglect these complex processes.

There are many sources of heating for interstellar dust. Among them, the ultraviolet and visible light emitted by massive stars is the strongest and most significant part~\cite{1983A&A...128..212M}. High-energy particles can also heat up dust through collisions, the interaction with the cosmic microwave background and the frictional collisions with gases~\cite{2022ApJS..263....5K,2015ApJ...805...59I,2013ApJ...766...13D,2001ApJ...554..778L,2017A&A...604A..58H}. The heating sources considered in this work are PBHs, which heat the dust through Hawking radiation.  Here we adopt the MRN model for the dust distribution~\footnote{Other distribution model can also be found, e.g., in Ref.~\cite{2025MNRAS.543.1574G}.}. In this model, interstellar dust is treated as a mixture of spherical graphite and silicate grains. Dust heats up after absorbing photons and simultaneously cools. The cooling rate of the dust is given by~\cite{2023PhRvD.107f3535M}:

\begin{equation}
\frac{\mathrm{d}E^{\mathrm{rad}}}{\mathrm{d}t}=4\pi \sigma_{\mathrm{d}} \int_{0}^{\infty} Q(\lambda)\,B(T_{\mathrm{d}},\lambda)\,\mathrm{d}\lambda,
\end{equation}
where $\sigma_{\mathrm{d}}$ is the cross-sectional area of dust particles and $Q(\lambda)$ is the absorption efficiency with wavelength $\lambda$~\cite{2023PhRvD.107f3535M}$:$
\begin{equation}
Q(\lambda) =
\begin{cases}
1, & \lambda \le 2\pi a \\[4pt]
\dfrac{2\pi a}{\lambda}, & \lambda > 2\pi a.
\end{cases}
\end{equation}
$B(T_{\mathrm{d}},\lambda)$ is the Planck function~\cite{2005pcim.book.....T}$:$
\begin{equation}
B(T_{\mathrm{d}},\lambda)=\frac{2hc^2}{\lambda^5} \frac{1}{\exp\left(\frac{hc}{\lambda\,k_{\mathrm{B}}T_{\mathrm{d}}}\right)-1},
\end{equation}
where $T_{\mathrm{d}}$ is the equilibrium temperature. Following Refs.~\cite{2005pcim.book.....T,Melikhov:2022mav,2023PhRvD.107f3535M}, we adopt the following expressions to calculate the equilibrium temperature:

\begin{equation}
T^{\mathrm{sil}}_{\mathrm{d}} = 13.6\left(\frac{1\,\mathrm{\mu m}}{a}\right)^{0.06}\,\mathrm{K},
\label{eq:cool_rate_sil}
\end{equation}

\begin{equation}
T^{\mathrm{gra}}_{\mathrm{d}} = 15.8\left(\frac{1\,\mathrm{\mu m}}{a}\right)^{0.06}\,\mathrm{K}.
\label{eq:cool_rate_gra}
\end{equation}

Given that large grains maintain equilibrium temperatures while small grains exhibit transient temperature spikes, we here consider 
dust grains with radius $a\gtrsim 0.01\mu \rm m$. Substituting $a=0.01\mu \rm m$ yields the conservative equilibrium temperatures: 
$T^{\mathrm{sil}}_{\mathrm{d}} = 17.93$ and $\mathrm{K},\ T^{\mathrm{gra}}_{\mathrm{d}}= 20.83\,\mathrm{K}$. Substituting these 
equilibrium temperatures into the above formulas gives the cooling rates of two different types of dust components. 
For the silicate component, we have 
\begin{equation}
\frac{\mathrm{d}E^{\mathrm{rad}}_{\mathrm{sil}}}{\mathrm{d}t}=2.21\times 10^{-14}\,\rm erg/s,
\end{equation}
and for the graphite component, 
\begin{equation}
\frac{\mathrm{d}E^{\mathrm{rad}}_{\mathrm{gra}}}{\mathrm{d}t}=4.68\times 10^{-14}\,\rm erg/s.
\end{equation}

\section{The heating effect from PBHs}
\label{sec:heat_from_pbh}

\begin{figure}[htbp] 
\begin{minipage}{\columnwidth}
\centering
\includegraphics[width=\columnwidth]{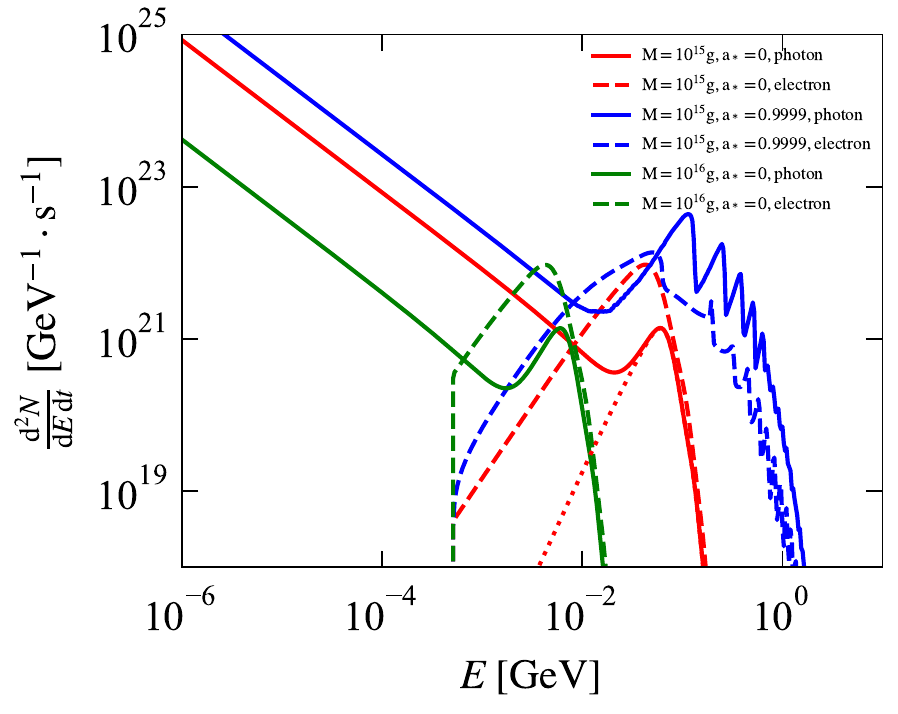}
\end{minipage}
\caption{Total photons spectra (solid lines) and electron spectra (dashed lines) emitted by PBHs with masses 
$M_{\rm PBH}=10^{15}\rm g$ and $10^{16}\rm g$, and spin parameter $a_{*}=0$ and 0.9999. The dotted red line corresponds to the primary 
photon spectrum for $M_{\rm PBH}=10^{15}\rm g$ with $a_{*}=0$, shown for comparison. The low-energy cutoff in the electron spectra corresponds 
to the electron rest mass, $m_e$ = 0.511 MeV. }
\label{fig:pbh_spectrum}
\end{figure}

PBHs can heat cosmic dust via photons emitted through Hawking radiation. On the one hand, the 'primary' photons are directly radiated by PBHs 
and subsequently heat the interstellar dust, as investigated in Ref.~\cite{2023PhRvD.107f3535M}. On the other hand, 'secondary' photons can also be produced during the Hawking radiation process. These originate from the decay of gauge bosons and from the decay of hadrons generated via 
the fragmentation of primary quarks and gluons emitted in the Hawking evaporation. Such secondary photons were not included in previous work. 
Including both primary and secondary components, the total photon spectrum $\frac{d N^{\rm tot}_{\gamma } }{dE dt}$ emitted by a PBH can be expressed as~\cite{2010PhRvD..81j4019C}
\begin{equation}
\frac{d N^{\rm tot}_{\gamma } }{dE dt}
=\frac{d N_{\gamma }^{\mathrm{pri}} }{d E dt}
+ \frac{d N_{\gamma }^{\mathrm{sec}} }{d E dt}.
\end{equation}
In this work, the total PBH photon spectrum $\frac{d N^{\rm tot}_{\gamma } }{dE dt}$ is calculated using the public code 
\texttt{BlackHawk\_v2.3}~\footnote{https://blackhawk.hepforge.org/}. The photon spectra for PBH masses $M_{\rm PBH}=10^{15}\rm g$ and $10^{16}\rm g$, 
with spin parameter $a_{*}=0$ and 0.9999, are shown in Fig.~\ref{fig:pbh_spectrum} (solid lines). It is evident that spinning PBHs emit significantly more photons than their non-spinning counterparts, while more massive PBHs produce fewer photons. The contribution from secondary photons is confined to energies below the spectral peak, becoming appreciable in the lower-energy range.

The radiation flux $F_{\rm PBH}$ from PBHs can be written as~\cite{2023PhRvD.107f3535M}
\begin{equation}
\begin{aligned}
F_{\mathrm{PBH}} &= f_{\rm PBH}\Omega_{\rm dm} \frac{n_{\mathrm{d}}}{N_{\mathrm{d}}}
\int_{0}^{R_{G}} \rho(r)\, r^{2}\,dr
\int_{0}^{\pi} \sin\theta \,d\theta
\int_{0}^{2\pi} d\varphi \\
&\quad \times \int_{0}^{R_{G}} r_{\mathrm{d}}^{2}\,dr_{\mathrm{d}}
\int_{0}^{\pi} \sin\theta_{\mathrm{d}} \,d\theta_{\mathrm{d}}
\int_{0}^{2\pi} \frac{1}{4\pi R^{2}} d\varphi_{\mathrm{d}} \\
&\quad \times \int_{M_{\min}}^{M_{\max}} \frac{\psi(M)}{M}\,dM
\int_{0}^{\infty} E \,\frac{dN^{\rm tot}_{\gamma}}{dE dt}\,dE,
\end{aligned}
\label{eq:flux_pbh}
\end{equation}
where $f_{\rm PBH}=\Omega_{\rm PBH}/\Omega_{\rm dm}$ denotes the mass fraction of PBHs in dark matter. Here, $\Omega_{i}=\rho_{i}/\rho_{\rm cri}$ 
is the dimensionless density parameter for component $i$, with $\rho_{\rm cri}$ being the critical density of the Universe. 
The quantities $n_d$ and $N_d$ represent the number density and total number of dust grains, respectively. 

The dark matter density profile in the Galaxy is denoted by $\rho(r)$ in Eq.~(\ref{eq:flux_pbh}). In addition to the classical 
Navarro-Frenk-White (NFW) model~\cite{Navarro:1995iw}, several other profiles have been proposed based on observational data. 
Unlike previous work, which only considered the NFW model, we here include four additional models for comparison: 
Einasto~\cite{Graham:2005xx,Navarro:2008kc}, Isothermal~\cite{10.1093/mnras/249.3.523,1980ApJS...44...73B}, 
Burkert~\cite{Burkert:1995yz}, and Moore~\cite{Diemand:2004wh}. The explicit forms of these dark matter density profiles are 
listed in Tab.~\ref{tab:dm_profile}, and the corresponding parameters $r_s$, $\rho_s$, and $\alpha$ for the Einasto model are given in Tab.~\ref{tab:dm_profile_para}. The quantity $R_{G}$ is the radius of the dark matter halo, determined via the relation $\int^{R_{G}}_{0} \rho(r)4\pi r^2 dr =M_{\rm MW,dm}$, where $M_{\rm MW,dm}$ is the dark matter mass of the Milky Way. Various estimates of the Milky Way's mass exist in the literature, with values typically ranging from $5\times 10^{11} M_{\odot}$ to $20\times 10^{11}\,M_{\odot}$~\cite{2025Univ...11..144M}. 
In this work, we adopt a total Milky Way mass of $M_{\rm MW} =10^{12}\,M_{\odot}$, and assume that dark matter constitutes 95\% of 
the Galactic mass~\cite{2017MNRAS.465...76M}, i.e., $M_{\rm MW,dm}=0.95M_{\rm MW}$. The resulting values of $R_G$ for each profile are 
summarized in Tab.~\ref{tab:dm_profile_para}.

The quantity $R$ in Eq.~(\ref{eq:flux_pbh}) denotes the distance between a PBH and a dust particle, 
\begin{equation}
\begin{aligned}
R^{2} = 
&\bigl(r\sin\theta\cos\varphi - r_{\mathrm{d}} \sin\theta_{\mathrm{d}} \cos\varphi_{\mathrm{d}}\bigr)^{2} \\
&+\bigl(r\sin\theta\sin\varphi - r_{\mathrm{d}} \sin\theta_{\mathrm{d}} \sin\varphi_{\mathrm{d}}\bigr)^{2} \\
&+\bigl(r\cos\theta - r_{\mathrm{d}} \cos\theta_{\mathrm{d}}\bigr)^{2},
\end{aligned}
\end{equation}
where $(r,\theta,\varphi)$ and $(r_{\mathrm{d}},\theta_{\mathrm{d}},\varphi_{\mathrm{d}})$ are the coordinates of the PBH and 
the dust particle, respectively, in a spherical coordinate system. 
In the above equation, we have assumed that the dust is uniformly distributed, following Ref.~\cite{2023PhRvD.107f3535M}.

The PBH mass function $\psi(M)$ is often idealized as monochromatic, i.e., $\sim \delta(M-M_{\rm PBH})$. However, in most inflationary scenarios, a more physical extended mass function is expected, commonly modeled as a lognormal distribution~\cite{PhysRevD.47.4244,Clesse:2015wea,Green:2016xgy,Dolgov:2008wu,Kannike:2017bxn}:

\begin{equation}
\psi(M) = \frac{1}{\sqrt{2\pi}\sigma M} \exp\left[-\frac{\ln^2(M/M_c)}{2\sigma^2}\right],
\label{eq:pbh_log_dis}
\end{equation}
where $\sigma$ denotes the width of the spectrum, and $M_c$ is the characteristic mass, corresponding to the peak value of mass fraction. This form accommodates various inflationary predictions and provides a realistic framework for observational constraints. In this work, we consider both of these mass functions, and our results can be straightforwardly extended to other forms 
as well~\cite{2020ARNPS..70..355C,2017PhRvD..96b3514C}.

The heating rate of dust by PBH Hawking radiation can be written as~\cite{2023PhRvD.107f3535M}
\begin{equation}
\frac{\mathrm{d}E^{\mathrm{abs}}}{\mathrm{d}t}
=4\pi \sigma_{\mathrm{d}} F_{\mathrm{PBH}}.
\label{eq:heating_rate}
\end{equation}

\begin{table}
\centering
\caption{Functional forms of the dark matter halo models used in this work. The corresponding model parameters are listed in Tab.~\ref{tab:dm_profile_para}.}

\begin{tabular*}{\columnwidth}{@{\extracolsep{\fill}}ll@{}}
\hline
\textbf{DM halo} & \textbf{Function form}  \\
\hline
NFW        & $\rho(r)=\rho_{s} \frac{r_{s}}{r}\left(1+\frac{r}{r_{s}}\right)^{-2}$ \\
Einasto    & $\rho(r)=\rho_{s}\exp\left\{ -\frac{2}{\alpha}\left[\left(\frac{r}{r_{s}}\right)^{\alpha}-1\right] \right\}$ \\
Isothermal & $\rho(r)=\frac{\rho_{s}}{1+\left(\frac{r}{r_{s}}\right)^{2}}$ \\
Burkert    & $\rho(r)=\frac{\rho_{s}}{\left(1+\frac{r}{r_{s}}\right)\left[1+\left(\frac{r}{r_{s}}\right)^{2}\right]}$ \\
Moore      & $\rho(r)=\rho_{s}\left(\frac{r_{s}}{r}\right)^{1.16}\left(1+\frac{r}{r_{s}}\right)^{-1.84}$ \\
\hline
\end{tabular*}

\label{tab:dm_profile}
\end{table}

\begin{table}
\centering
\caption{Parameter values for the dark matter halo models shown in Tab.~\ref{tab:dm_profile}. The values of $r_s$, $\rho_s$, and $\alpha$ are taken from Ref.~\cite{2011JCAP...03..051C}. $R_G$ denotes the radius of the dark matter halo of the Milky Way; the procedure for determining $R_G$ is described in the main text.}

\begin{tabular*}{\columnwidth}{@{\extracolsep{\fill}}lllll@{}}
\hline
DM halo & $\alpha$ & $r_{s}\,[\mathrm{kpc}]$ & $\rho_{s}\,[\mathrm{GeV}/\mathrm{cm}^{3}]$ & $R_{G}\,[\mathrm{kpc}]$ \\
\hline
NFW        & $-$    & 24.42 & 0.184 & 141.8 \\
Einasto    & 0.17   & 28.44 & 0.033 & 128.8 \\
Isothermal & $-$    & 4.38  & 1.387 & 113.9  \\
Burkert    & $-$    & 12.67 & 0.712 & 186.1 \\
Moore      & $-$    & 30.28 & 0.105 & 138.5 \\
\hline
\end{tabular*}
\label{tab:dm_profile_para}
\end{table}


In addition to photons, PBHs also emit electrons via Hawking radiation. 
To assess their contribution to dust heating, we consider the energy 
deposition efficiency of electrons in dust grains~\cite{1986ApJ...302..363D}, 
and more importantly, the finite range of electrons in the interstellar 
medium. Unlike photons, which can be absorbed and scattered by dust 
but otherwise propagate freely through the interstellar medium (ISM) to heat distant grains, 
electrons lose energy continuously through inelastic collisions and 
are severely limited in their propagation distance. For MeV-scale 
electrons, which dominate the Hawking spectrum for PBH masses in the 
$10^{15}$--$10^{17}\,\mathrm{g}$ range (see Fig.~\ref{fig:pbh_spectrum}), the stopping range in the ISM 
is only $R_e \sim 0.1$ kpc or less; even for 100 MeV electrons the range 
is at most $R_e \sim 1$ kpc in typical diffuse gas~\cite{PhysRev.73.449,1979ApJ...231...77D,2004ApJ...601..340K}
\footnote{Stopping Powers for Electrons and Positions, ICRU-37, 1984}. 
Since our dust distribution is assumed to be uniform throughout the 
Galactic halo of radius $R_G\sim 100$--180 kpc, only a tiny fraction 
of dust grains lie within the stopping distance of PBH-emitted 
electrons. The effective volume for electron heating scales as 
$(R_e/R_G)^3\lesssim 10^{-6}$, while the local heating efficiency 
enhancement partially compensates, leading to a total 
electron heating power that is at most of order a few percent of 
the photon heating power~\cite{2013A&A...556A...6B}. We therefore neglect 
electron heating in our main analysis; this approximation is 
conservative, as including electrons would only marginally strengthen 
the resulting constraints on $f_{\rm PBH}$.


\section{Constraints on the PBH dark matter fraction}
\label{sec:constraints}

After obtaining the cooling rates of the dust, $dE^{\rm rad}/dt$, as given by Eqs.~(\ref{eq:cool_rate_sil}) and (\ref{eq:cool_rate_gra}), 
and the heating rate from PBHs, $dE^{\rm heat}/dt$, as given by Eq.~(\ref{eq:heating_rate}), we derive the upper limits on the fraction of PBHs 
in dark matter by requiring that the heating rate does not exceed the cooling rate of the dust: 

\begin{equation}
\frac{dE^{\rm heat}}{dt} \leq \frac{dE^{\rm rad}}{dt}.	
\label{eq:limit}
\end{equation}

\begin{figure*}[t]
    \centering

    \begin{subfigure}[b]{0.32\textwidth}
        \centering
        \includegraphics[width=\linewidth]{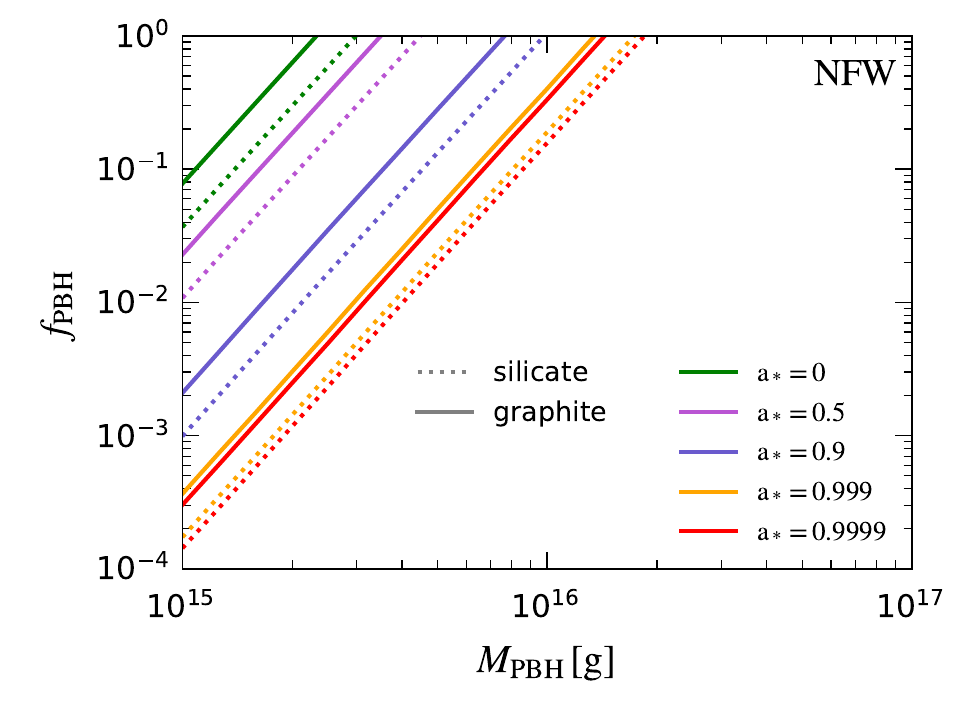}
        \label{fig:sub1}
    \end{subfigure}
    \hfill
    \begin{subfigure}[b]{0.32\textwidth}
        \centering
        \includegraphics[width=\linewidth]{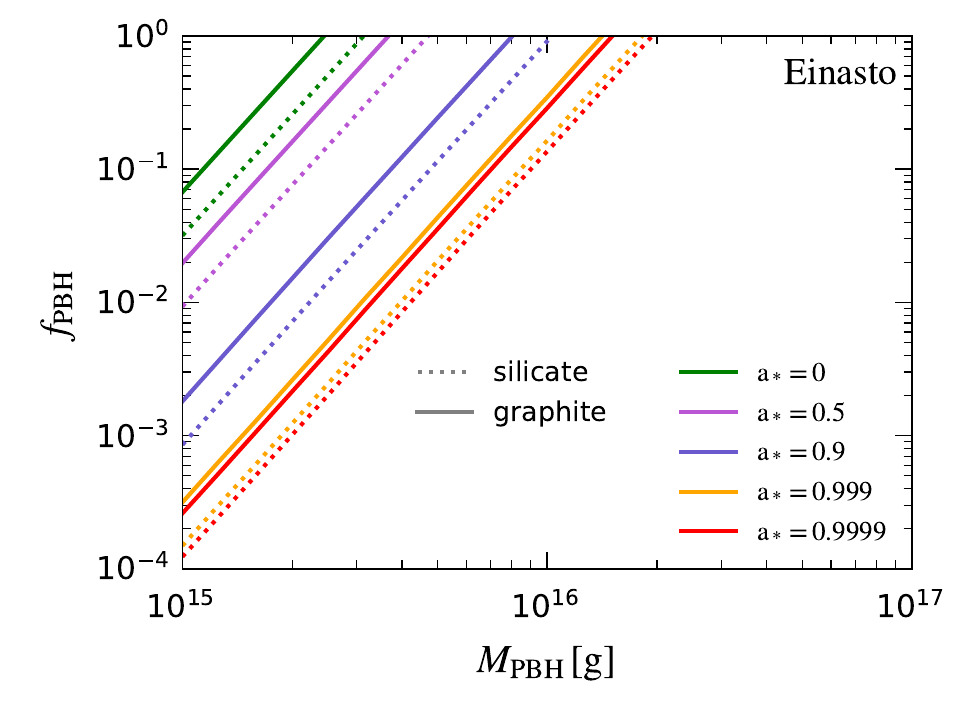}
        \label{fig:sub2}
    \end{subfigure}
    \hfill
    \begin{subfigure}[b]{0.32\textwidth}
        \centering
        \includegraphics[width=\linewidth]{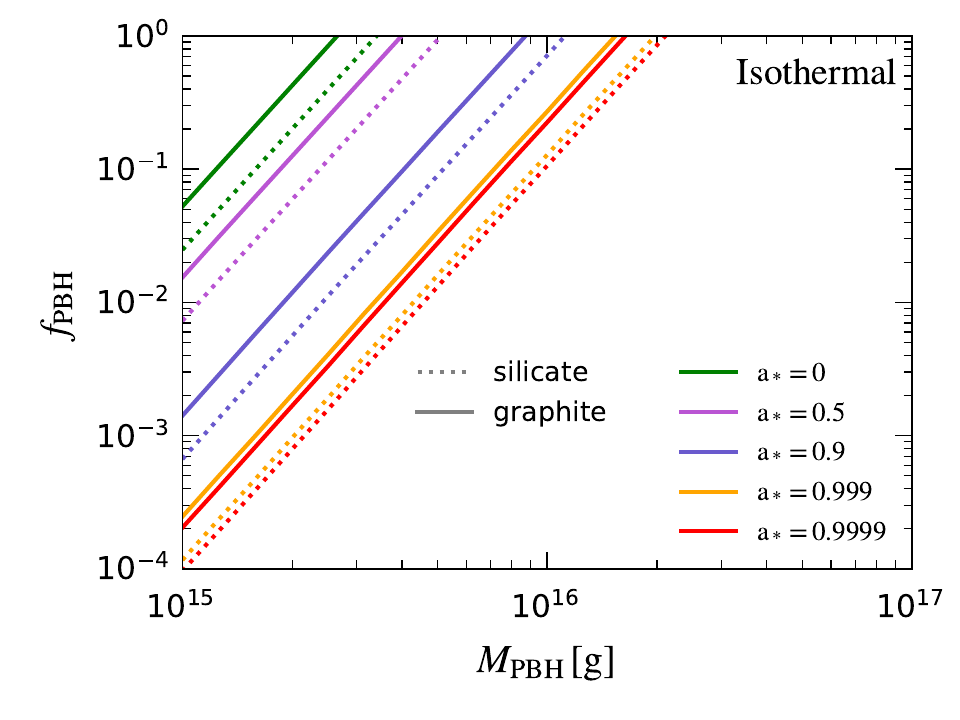}
        \label{fig:sub3}
    \end{subfigure}

    \begin{minipage}{1\textwidth}  
        \centering
        \begin{subfigure}[b]{0.32\linewidth} 
            \centering
            \includegraphics[width=\linewidth]{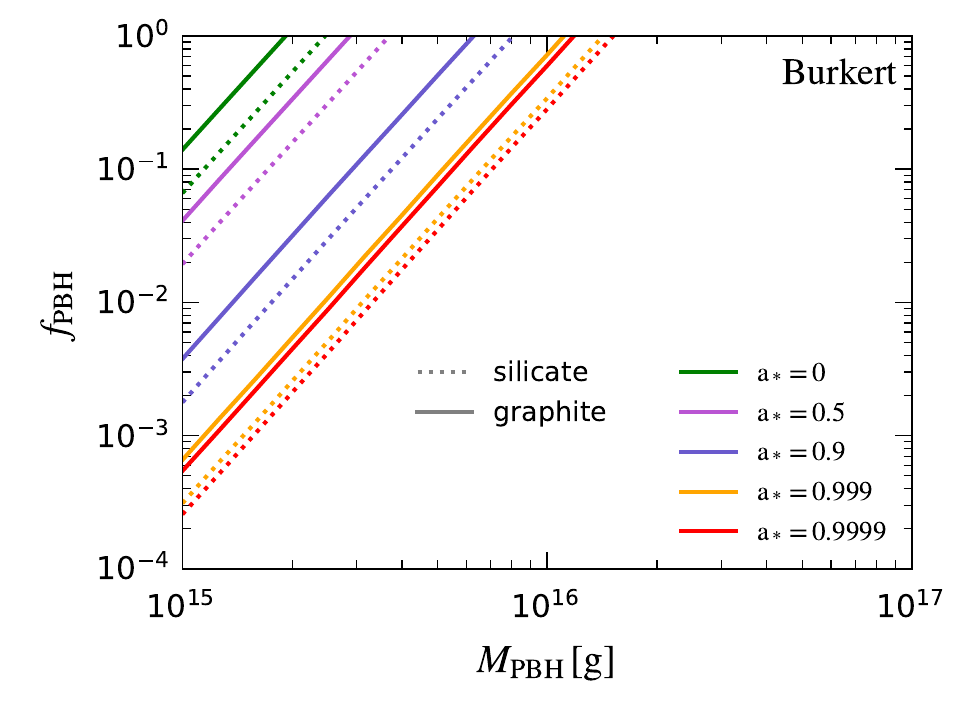}
            \label{fig:sub4}
        \end{subfigure}
        \hfill
        \begin{subfigure}[b]{0.32\linewidth}
            \centering
            \includegraphics[width=\linewidth]{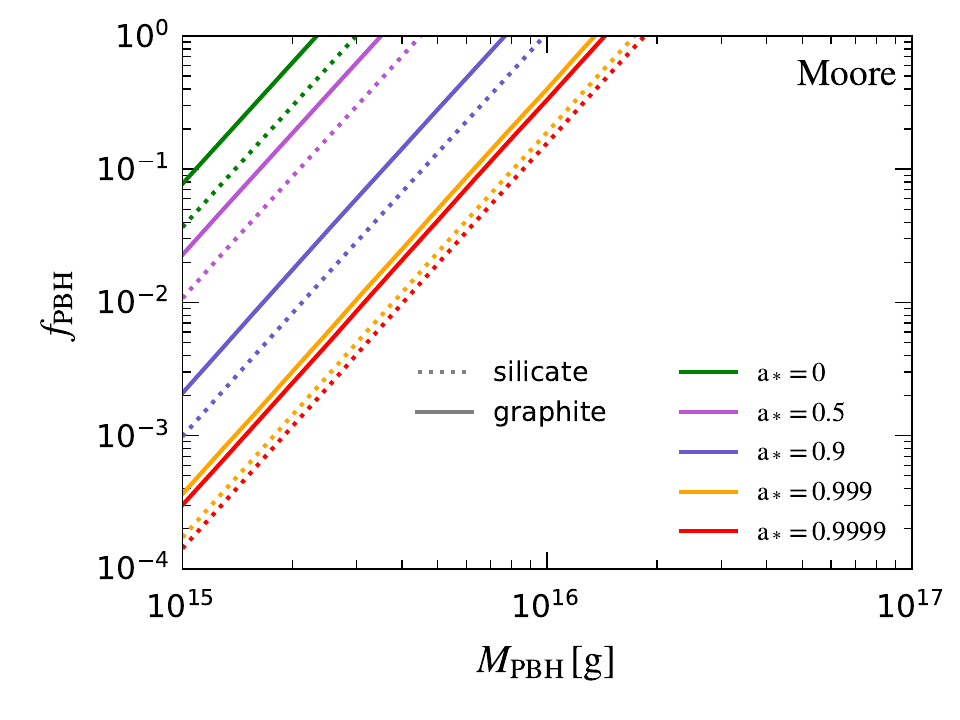}
            \label{fig:sub5}
        \end{subfigure}
        \hfill
         \begin{subfigure}[b]{0.32\linewidth}
            \centering
            \includegraphics[width=\linewidth]{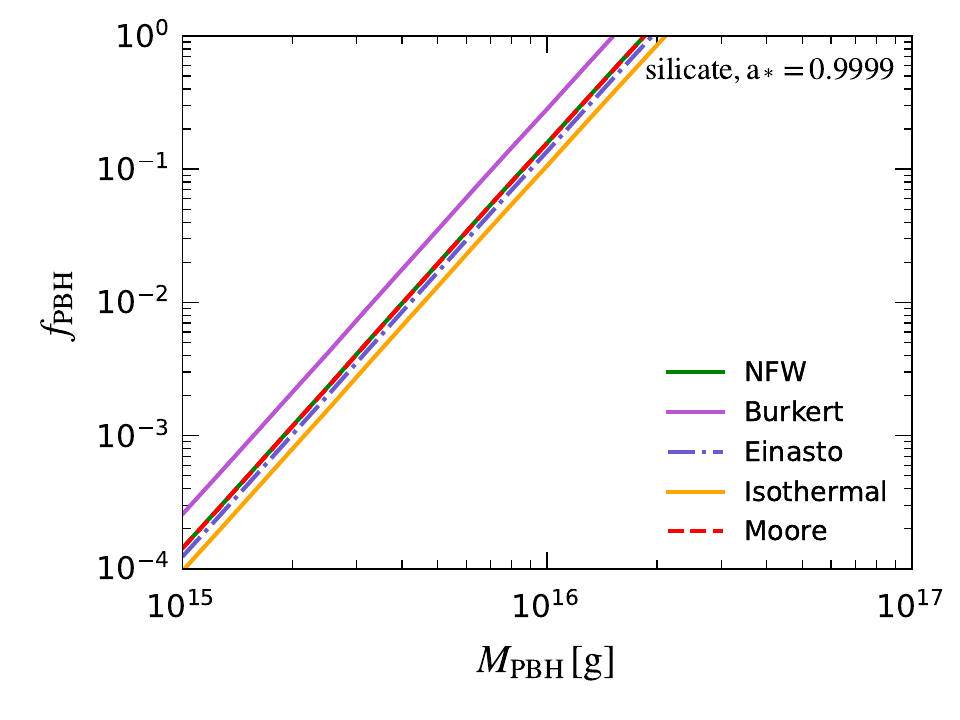}
            \label{fig:sub5}
        \end{subfigure}
    \end{minipage}

    \caption{Constraints on the PBH fraction $f_{\rm PBH}$ for different dark matter profiles (NFW, Einasto, Isothermal, Burkert, and Moore), 
    for silicate and graphite grains, and for spin parameters $a_{*}=0, 0.5,0.9, 0.999$ and 0.9999. All results are shown for a 
    monochromatic mass function. A comparison among the different dark matter profiles is also shown in the last subplot.} 
    \label{fig:constraints_delta}
\end{figure*}

The resulting constraints on the PBH dark matter fraction $f_{\rm PBH}$ for a monochromatic mass function are presented in Figs.~\ref{fig:constraints_delta}. 
These figures display the constraints for silicate and graphite grains separately, considering 
different dark matter density profiles and spin parameters $a_{*}=0,0.5,0.9,0.999$ and 0.9999. For a fixed spin parameter, 
the constraints on $f_{\rm PBH}$ are comparable across the different dark matter profiles. As expected, higher spin values lead to 
stronger constraints, owing to the increased photon emission associated with higher spin (as seen in Fig.~\ref{fig:pbh_spectrum}). 
Additionally, because silicate has a lower cooling rate than graphite, the constraints on $f_{\rm PBH}$ are stronger for silicate by 
a factor of $\sim 2.1$, reflecting the differences in the cooling rates given by Eqs.~(\ref{eq:cool_rate_sil}) and (\ref{eq:cool_rate_gra}). 
A comparison of the constraints for different dark matter profiles is also shown in Fig.~\ref{fig:constraints_delta} (last panel, for silicate 
with $a_{*}=0.9999$). Overall, the upper limits on $f_{\rm PBH}$ are of the order of $\sim 10^{-4}$ for all the dark matter profiles 
considered in this work. Specifically, the weakest limits are obtained for the Burkert profile, while the strongest limits 
arise from the Isothermal profile. The limits for the NFW and Moore profiles are consistent with each other 
and are stronger than those for the Einasto profile. Note that radiation from extragalactic PBHs can also heat the interstellar dust, and the corresponding 
constraints on $f_{\rm PBH}$ are comparable than those obtained in this work~\cite{Li:2026paw}.   


\begin{figure*}[t]
    \centering

    \begin{subfigure}[b]{0.32\textwidth}
        \centering
        \includegraphics[width=\linewidth]{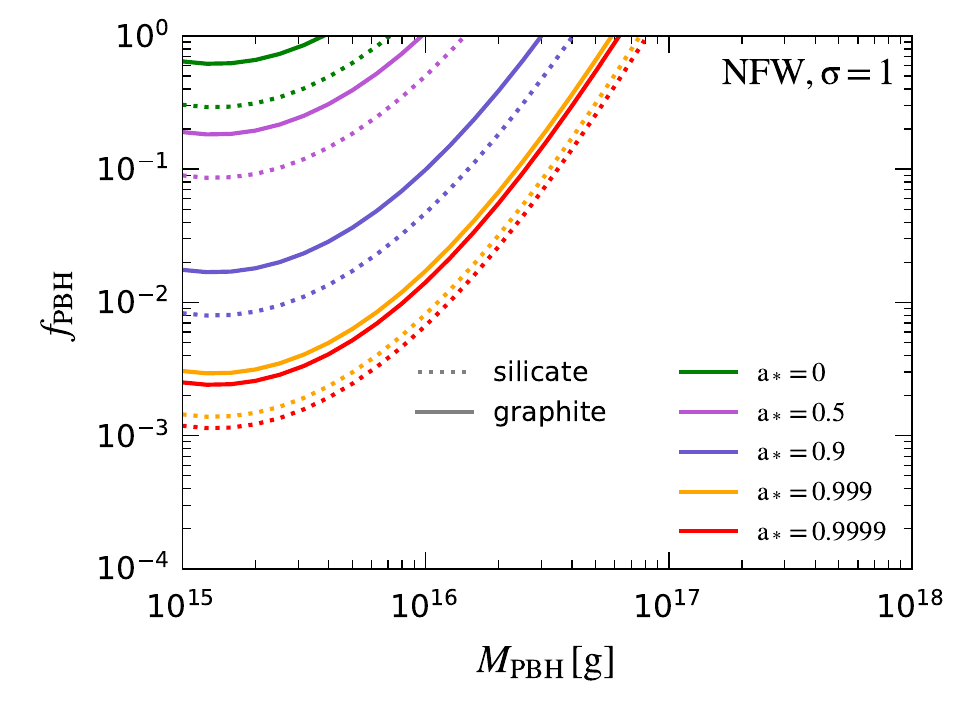}
        \label{fig:sub1}
    \end{subfigure}
    \hfill
    \begin{subfigure}[b]{0.32\textwidth}
        \centering
        \includegraphics[width=\linewidth]{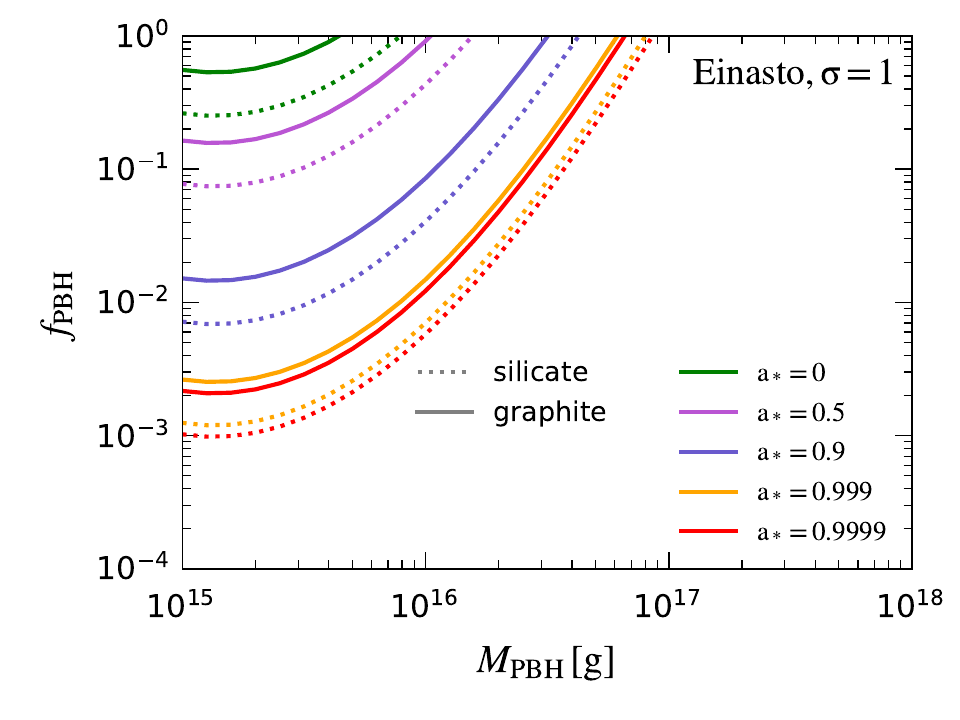}
        \label{fig:sub2}
    \end{subfigure}
    \hfill
    \begin{subfigure}[b]{0.32\textwidth}
        \centering
        \includegraphics[width=\linewidth]{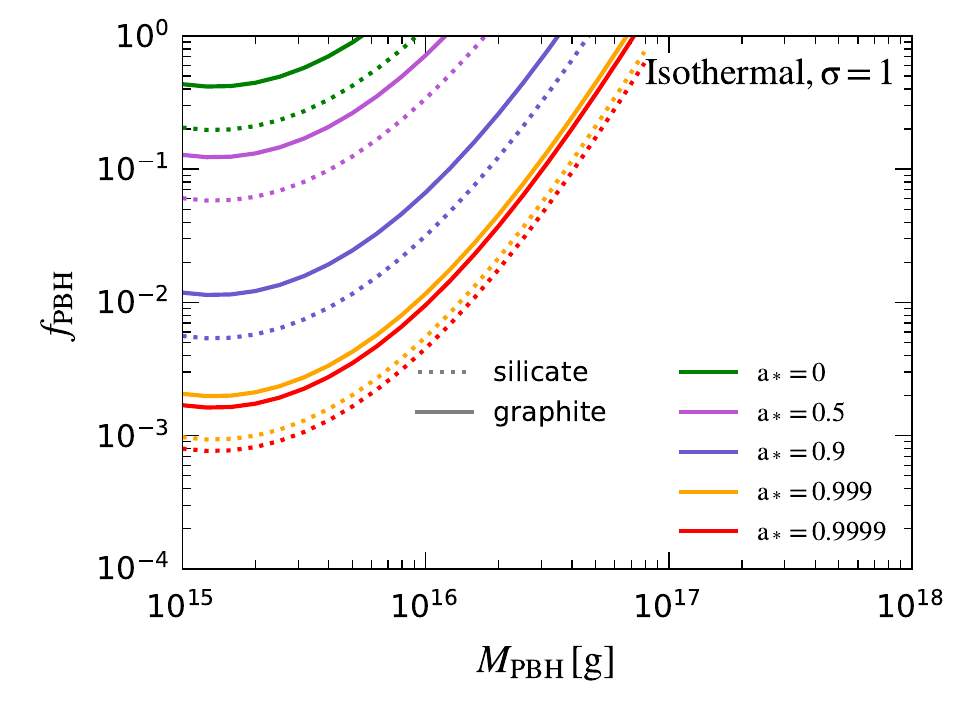}
        \label{fig:sub3}
    \end{subfigure}

    \begin{minipage}{1\textwidth}  
        \centering
        \begin{subfigure}[b]{0.32\linewidth} 
            \centering
            \includegraphics[width=\linewidth]{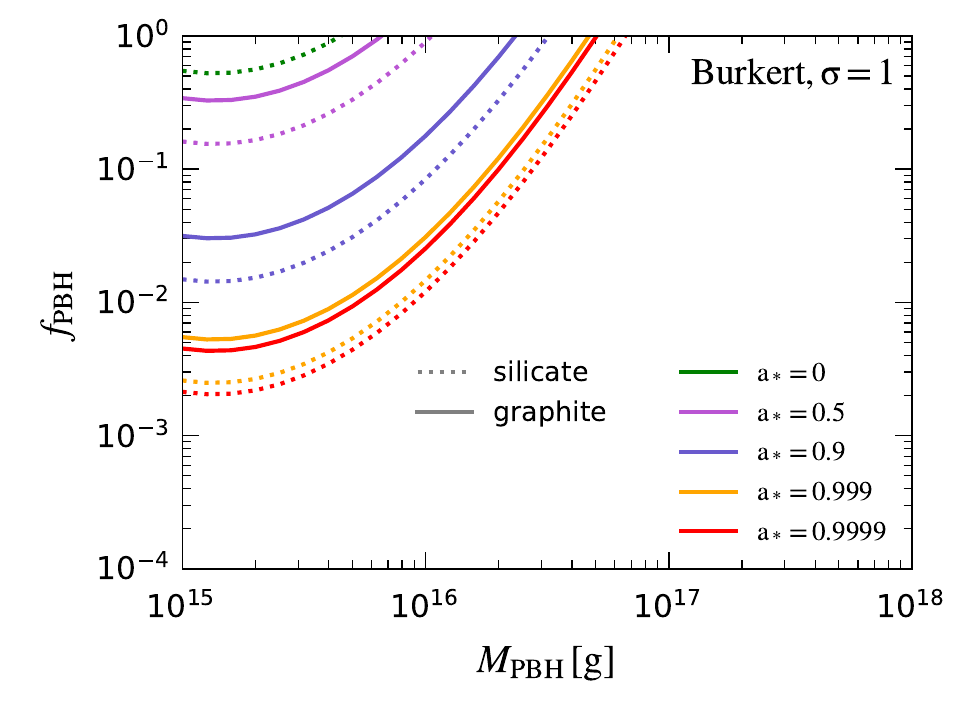}
            \label{fig:sub4}
        \end{subfigure}
        \hfill
        \begin{subfigure}[b]{0.32\linewidth}
            \centering
            \includegraphics[width=\linewidth]{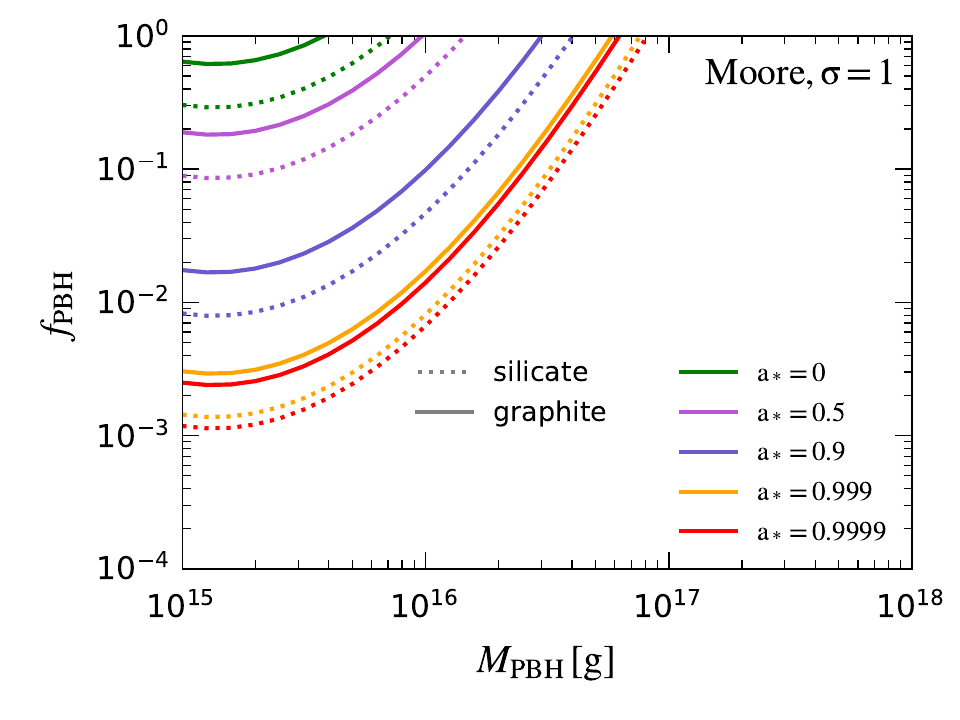}
            \label{fig:sub5}
        \end{subfigure}
        \hfill
         \begin{subfigure}[b]{0.32\linewidth}
            \centering
            \includegraphics[width=\linewidth]{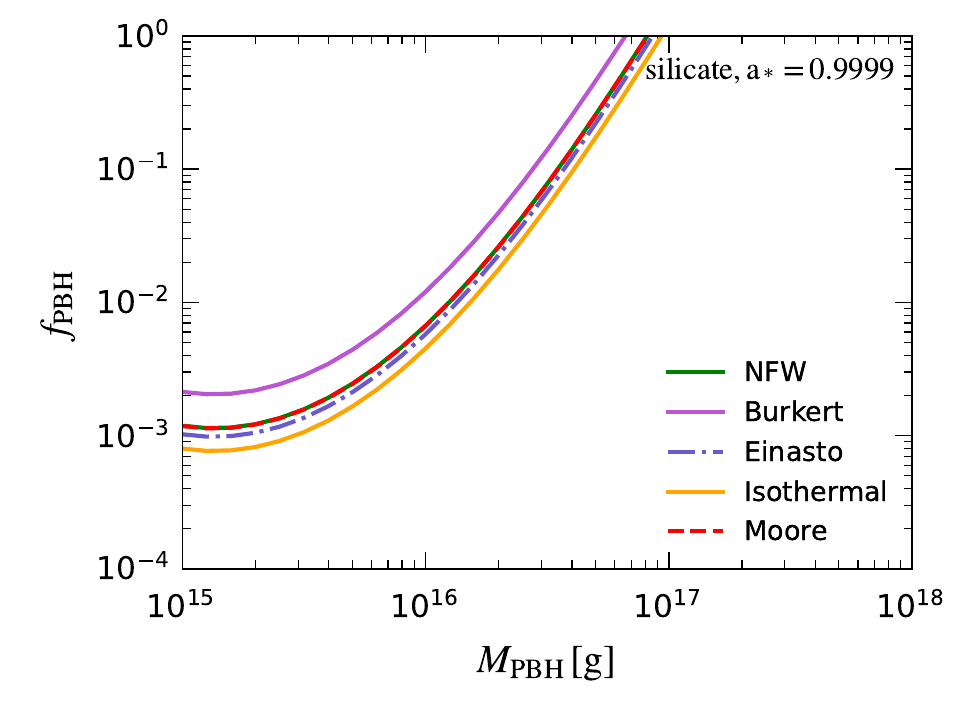}
            \label{fig:sub5}
        \end{subfigure}
    \end{minipage}

    \caption{Constraints on the PBH fraction $f_{\rm PBH}$ for different dark matter profiles (NFW, Einasto, Isothermal, Burkert, and Moore), 
    for silicate and graphite grains, and for spin parameters $a_{*}=0, 0.5,0.9, 0.999$ and 0.9999. All results are shown for 
    a lognormal mass function with $\sigma=1$. A comparison among the different dark matter profiles is also shown in the last subplot. 
    Note that $M_{\rm PBH}=M_c$ denotes the characteristic mass in Eq.~(\ref{eq:pbh_log_dis}).} 
    \label{fig:constraints_sigma_1}
\end{figure*}

\begin{figure*}[t]
    \centering

    \begin{subfigure}[b]{0.32\textwidth}
        \centering
        \includegraphics[width=\linewidth]{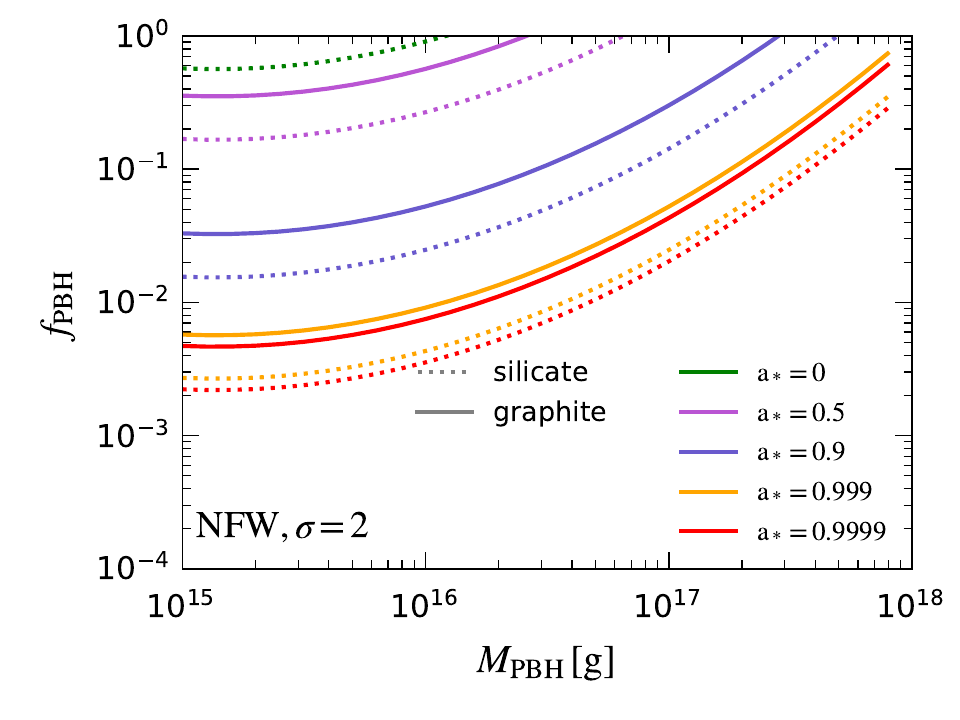}
        \label{fig:sub1}
    \end{subfigure}
    \hfill
    \begin{subfigure}[b]{0.32\textwidth}
        \centering
        \includegraphics[width=\linewidth]{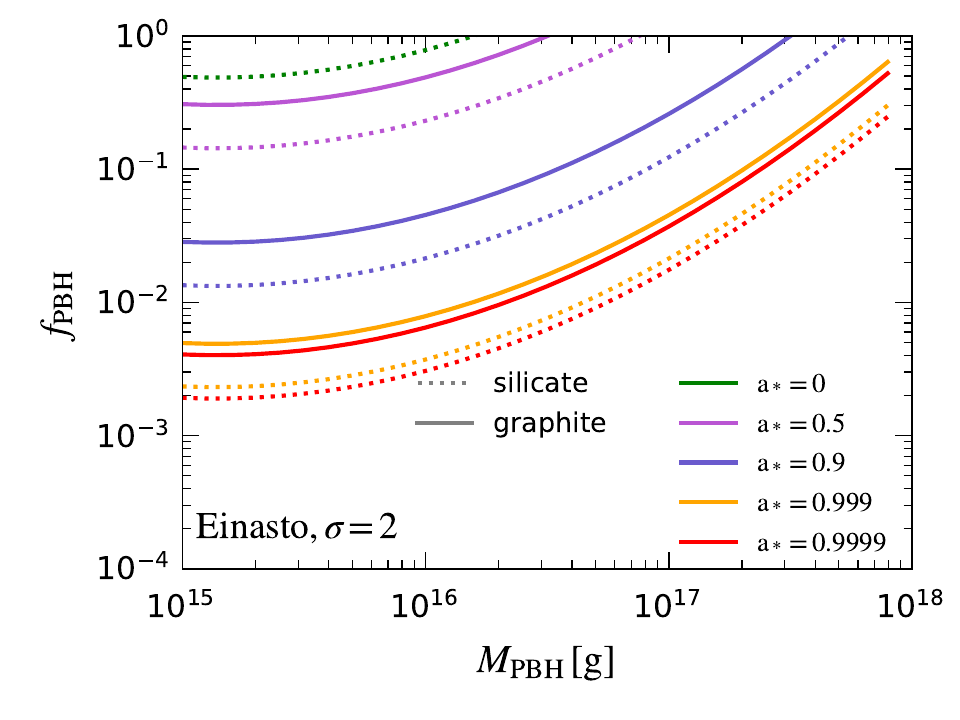}
        \label{fig:sub2}
    \end{subfigure}
    \hfill
    \begin{subfigure}[b]{0.32\textwidth}
        \centering
        \includegraphics[width=\linewidth]{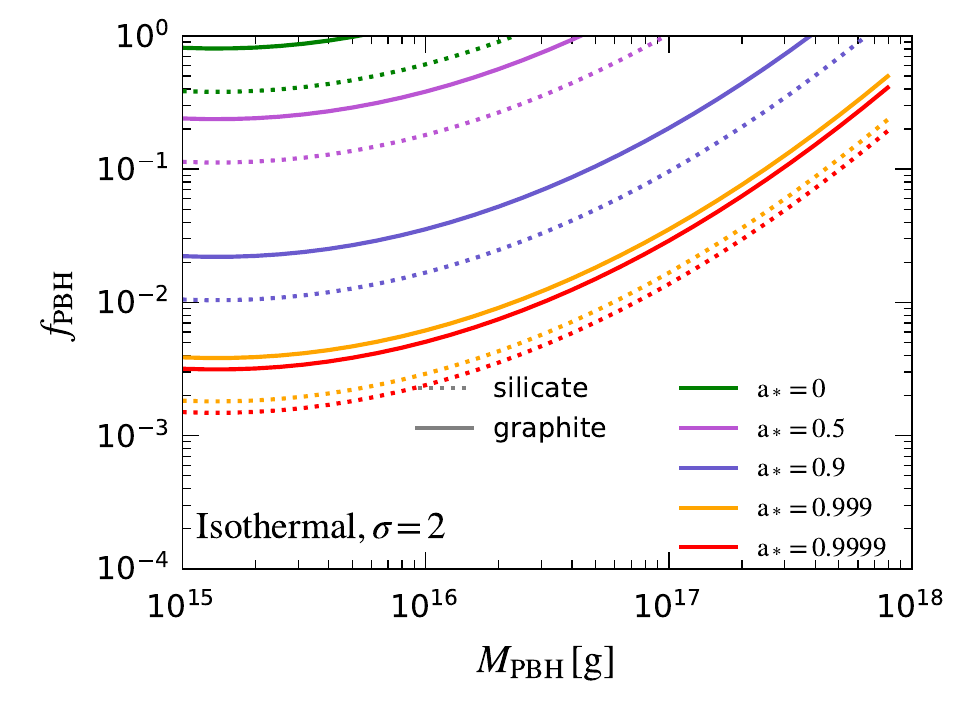}
        \label{fig:sub3}
    \end{subfigure}

    \begin{minipage}{1\textwidth}  
        \centering
        \begin{subfigure}[b]{0.32\linewidth} 
            \centering
            \includegraphics[width=\linewidth]{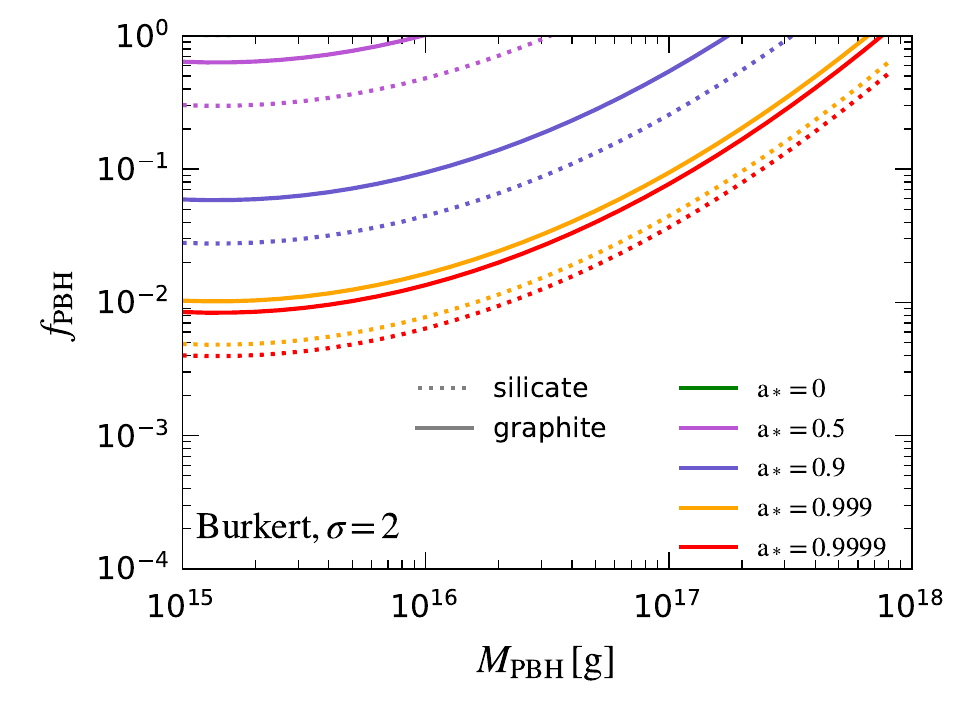}
            \label{fig:sub4}
        \end{subfigure}
        \hfill
        \begin{subfigure}[b]{0.32\linewidth}
            \centering
            \includegraphics[width=\linewidth]{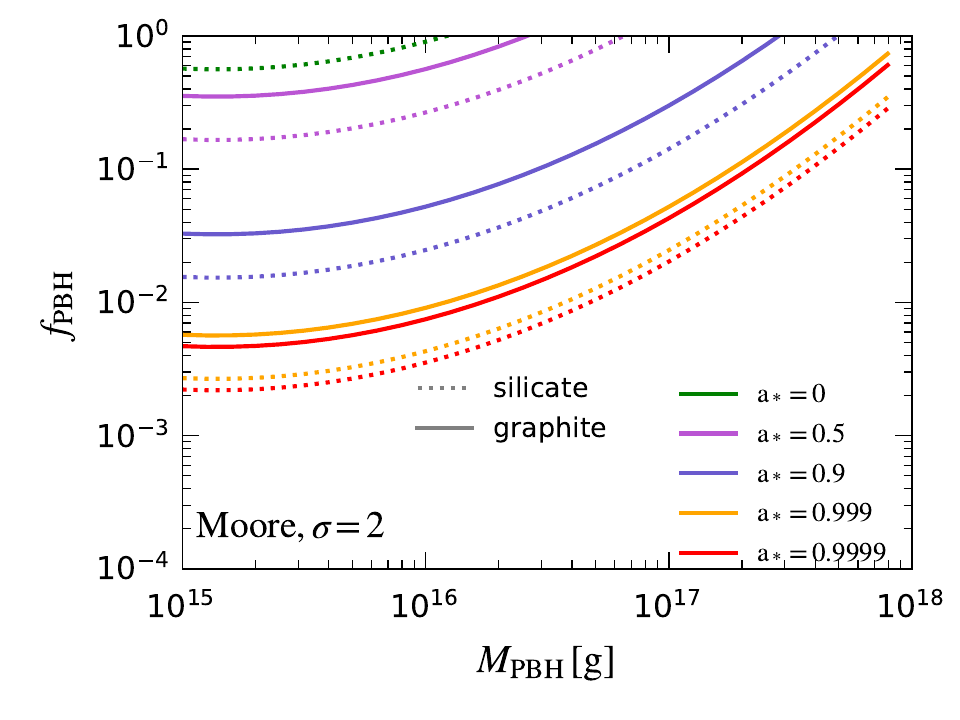}
            \label{fig:sub5}
        \end{subfigure}
        \hfill
         \begin{subfigure}[b]{0.32\linewidth}
            \centering
            \includegraphics[width=\linewidth]{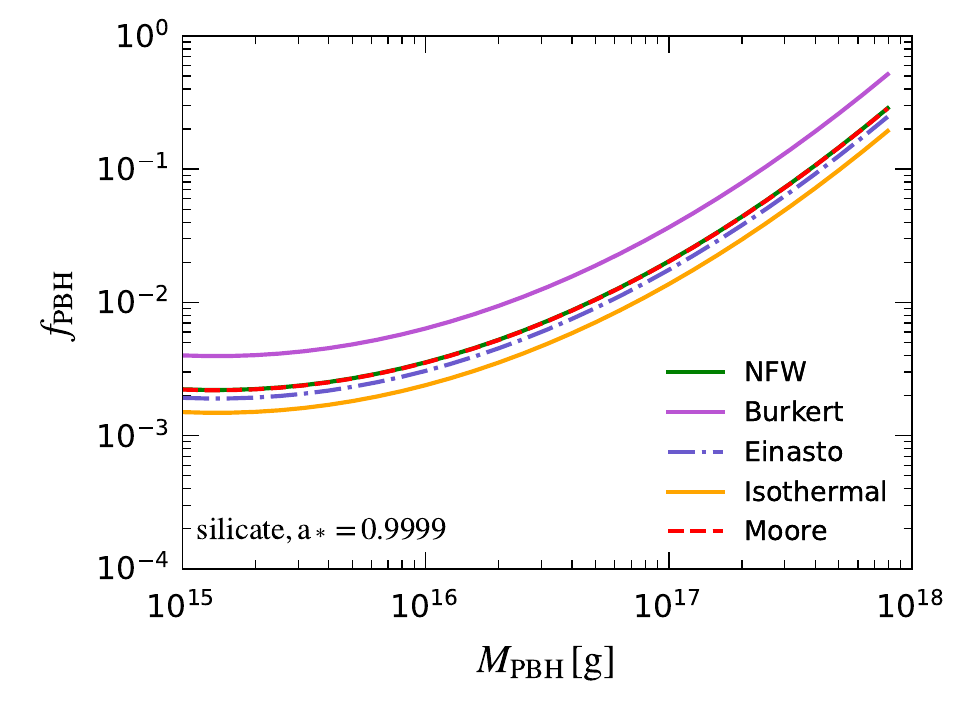}
            \label{fig:sub5}
        \end{subfigure}
    \end{minipage}

    \caption{Constraints on the PBH fraction $f_{\rm PBH}$ for different dark matter profiles (NFW, Einasto, Isothermal, Burkert, and Moore), 
    for silicate and graphite, and for spin parameters $a_{*}=0, 0.5,0.9, 0.999$ and 0.9999. All resluts are shown for a lognormal mass function with 
    $\sigma=2$. 
    A comparison among the different dark matter profiles is also shown in the last subplot. Note that $M_{\rm PBH}=M_c$ here denotes 
    the characteristic mass in Eq.~(\ref{eq:pbh_log_dis}).} 
    \label{fig:constraints_sigma_2}
\end{figure*}

The constraints on $f_{\rm PBH}$ for a lognormal mass function of PBHs are shown in Figs.~\ref{fig:constraints_sigma_1} ($\sigma=1$) and~\ref{fig:constraints_sigma_2} ($\sigma=2$), respectively. The overall trend of the limits is consistent with that obtained for the monochromatic mass function. 
It can be seen that the constraints become weaker for smaller PBH masses, while for larger PBH masses, they exclude the possibility 
that PBHs constitute all of the dark matter, with this exclusion becoming more pronounced as $\sigma$ increases.


\begin{figure}
\centering
\includegraphics[width=0.5\textwidth]{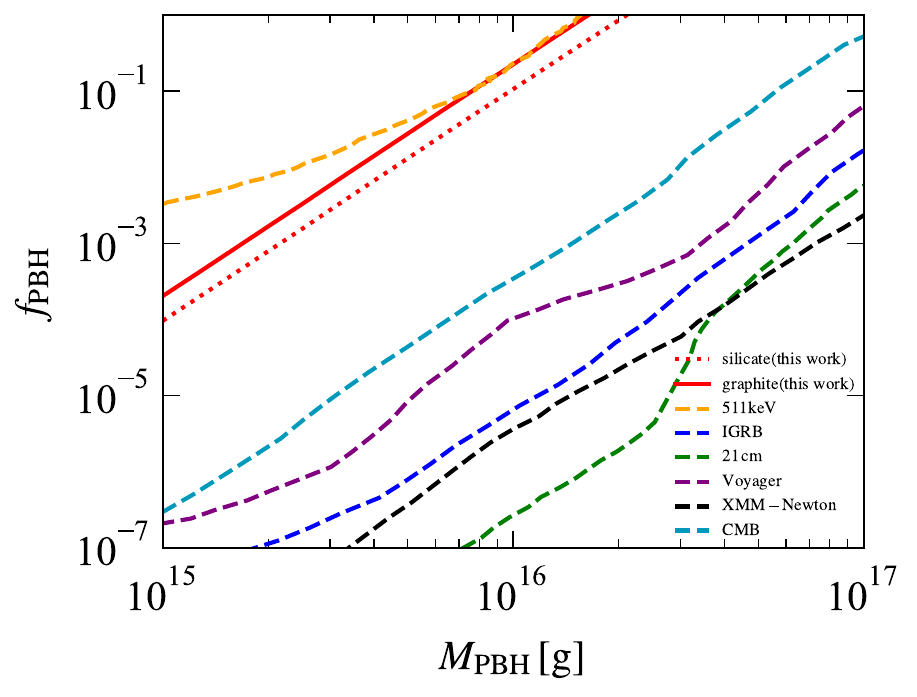}
\caption{Comparison of constraints on spinning PBHs with several other existing limits: 
(1) from $511\,\mathrm{keV}$ positron annihilation at the Galactic center (marked as "511keV")~\cite{2020PhRvL.125j1101D}; 
(2) from the isotropic gamma-ray background (marked as "IGRB")~\cite{2020PhRvD.101b3010A}; 
(3) from the global 21 cm differential brightness temperature (marked as "21cm")~\cite{2022MNRAS.510.4236N}; 
(4) from the measurements of positrons and electrons from the Voyager spacecraft (marked as "Voyager")~\cite{2024PhRvD.110l3022L}; 
(5) from X-ray diffuse observations using the XMM-Newton data (marked as "XMM-Newton")~\cite{2024PhRvD.110l3022L}; 
(6) from the studies of the effects of PBH radiation on the cosmic microwave background (Planck data) (marked as "CMB") \cite{2022JCAP...03..012C}. 
All limits are shown for spin parameter $a_{*}=0.9999$, except for the CMB results which use $a_{*}=0.999$. Our strongest limits 
(Isothermal profile, $a_{*}=0.9999$) are included for comparison. }
\label{fig:com_spin_9999}
\end{figure}

For comparison, Fig.~\ref{fig:com_spin_9999} shows the constraints on $f_{\rm PBH}$ for spinning PBHs from several existing limits:
(1) constraints from the $511\,\mathrm{keV}$ signals produced by positron annihilation at the Galactic center (labeled "511\,keV")~\cite{2020PhRvL.125j1101D}, noting that updated and stronger limits have been obtained in Ref.~\cite{2024PhRvD.110l3022L};
(2) constraints from the isotropic gamma-ray background (labeled "IGRB")~\cite{2020PhRvD.101b3010A};
(3) constraints from the global 21 cm differential brightness temperature (labeled "21\,cm")~\cite{2022MNRAS.510.4236N};
(4) constraints from Voyager measurements of positrons and electrons (labeled "Voyager")~\cite{2024PhRvD.110l3022L};
(5) constraints from diffuse X-ray observations with XMM-Newton (labeled "XMM-Newton")~\cite{2024PhRvD.110l3022L};
(6) constraints from the effects of PBH radiation on the cosmic microwave background (Planck data, labeled "CMB")~\cite{2022JCAP...03..012C}.

All limits are shown in Fig.~\ref{fig:com_spin_9999} for spin parameter $a_{*}=0.9999$, except for the CMB results, which adopt $a_{*}=0.999$; the difference in $f_{\rm PBH}$ limits between these two spin values should be small. For our constraints, we display the strongest limits obtained in this work, corresponding to the Isothermal profile with $a_{*}=0.9999$. As can be seen from the figure, the $511\,\mathrm{keV}$ constraint is comparable to our graphite limit for PBH masses $M_{\rm PBH}>6\times 10^{15}\,\mathrm{g}$, while our limits are stronger for smaller masses. All other existing constraints are stronger than our results.

Our analysis assumes a uniform dust distribution throughout the Galactic 
halo. In reality, interstellar dust is strongly concentrated in the 
Galactic disk. Recent high-precision extinction measurements reveal 
that the Milky Way's dust consists of two distinct components: a thin 
disk with a scale height of $\sim 80$ pc and a thick disk with a scale 
height of $\sim 150$ pc~\cite{2025MNRAS.543.1574G}. If we adopt a more realistic 
disk-like distribution, the dust density in the central region would be 
enhanced by a factor of $\sim 10^5$ relative to our uniform model, while 
only $\sim 15\%$ of the dark matter mass resides within the disk volume 
~\cite{2017MNRAS.465...76M}. This would increase the local heating rate by 
$\sim 10^4$, thereby tightening the upper limit on $f_{\rm PBH}$ in the 
disk by roughly four orders of magnitude, reaching $\mathcal{O}(10^{-8})$. 
Conversely, constraints in the halo would become significantly weaker. 
Thus, our uniform assumption yields a conservative, sky-averaged 
constraint, while realistic dust distributions would introduce strong 
directional dependence. A detailed investigation incorporating a 
realistic three-dimensional dust distribution is deferred to future work.

     
\section{Conclusion}
\label{sec:con}

We have investigated the heating of the interstellar dust by Hawking radiation from cosmologically distributed primordial black holes (PBHs) 
and derived updated constraints on the PBH dark matter fraction $f_{\rm PBH}$. Compared with previous works, our analysis incorporates 
several key improvements. We include secondary photons emitted during the Hawking evaporation process, which were neglected in earlier work. We also 
examine the impact of five different dark matter density profiles, namely NFW, Einasto, Isothermal, Burkert, and Moore, 
on the resulting constraints, thereby assessing the model dependence of the limits. More importantly, 
we have extended the analysis to spinning PBHs, which emit significantly more photons 
than their non-spinning counterparts. Furthermore, we consider both monochromatic and lognormal mass functions for PBHs.

Our results show that the constraints on $f_{\rm PBH}$ become considerably stronger as the spin parameter $a_{*}$ increases. 
For a fixed dark matter profile, the strongest upper limits are obtained for silicate grains with $a_{*}=0.9999$, reaching $f_{\rm PBH}\sim 10^{-4}$, 
owing to the lower cooling rate of silicate compared with graphite. Among the five dark matter profiles considered, the Isothermal model 
yields the strongest constraints, while the Burkert model gives the weakest. The limits derived from the NFW and Moore profiles 
are comparable to each other and are stronger than 
those from the Einasto profile. For all profiles, the strongest constraints are of the order of $\sim 10^{-4}$ for the 
monochromatic mass function. Similar trends are observed for the lognormal mass function, 
though the limits are weaker for smaller PBH masses and stronger for larger masses. Moreover, as $\sigma$ increases, 
a broader mass range is excluded for PBHs that would otherwise constitute all of the dark matter for the monochromatic mass function.

We have also compared our results with other existing constraints on PBHs. For our strongest constraints, 
corresponding to the Isothermal profile with $a_{*}=0.9999$, we find that our limits are stronger than those derived from 
the 511 keV positron annihilation signals at the Galactic center for PBH masses of $M_{\rm PBH}<6\times 10^{15}\rm g$. Although our constraints 
are weaker than most other existing bounds, the interstellar dust heating method provides a complementary and independent approach 
to probing the PBH parameter space.

\section*{Acknowledgements}

This work is supported by the Shandong Provincial Natural Science Foundation (Grant Nos. ZR2025MS16, ZR2025MS47, ZR2025QC25).


\bibliographystyle{apsrev4-1}
\bibliography{ref}


\end{document}